\documentclass[superscriptaddress,showpacs,amsmath,amssymb,pra,twocolumn]{revtex4-1}

\usepackage{tikz}
\usetikzlibrary{positioning}
\tikzset{>=stealth}

\usepackage{amsfonts}
\usepackage[english]{babel}
\usepackage{braket}

\usepackage[bookmarks=false,breaklinks,bookmarksopen,bookmarksnumbered,colorlinks,linkcolor=LINKCOL,linktocpage,citecolor=CITECOL,urlcolor=URLCOL,pdfpagemode=UseOutline,pdftex]{hyperref}

\usepackage{booktabs}
\usepackage{comment}

\definecolor{TITLECOL}{rgb}{0.1,0.2,0.7} 
\definecolor{SECOL}{rgb}{0.1,0.2,0.7} 
\definecolor{CONTENTSCOL}{rgb}{0.1,0.2,0.7} 
\definecolor{SSECOL}{rgb}{0.25,0,0.48} 
\definecolor{SSSECOL}{rgb}{0.2,0.08,0.53} 
\definecolor{FINCOL}{rgb}{0.01,0.3,0.07} 

\def\coloredauthor#1{\author{\textcolor{CITECOL}{#1}}} 

\definecolor{URLCOL}{rgb}{0,0.17,0.43} 
\definecolor{LINKCOL}{rgb}{0.05,0.4,0} 
\definecolor{CITECOL}{rgb}{0.35,0,0.48} 

\usepackage{natbib}
\def\bea{\begin{eqnarray}}
\def\eea{\end{eqnarray}}
\def\ben{\begin{equation}}
\def\een{\end{equation}}
\def\benu{\begin{enumerate}}
\def\enu{\end{enumerate}}

\def\bei{\begin{itemize}}
\def\eei{\end{itemize}}
\def\beit{\begin{itemize}}
\def\eit{\end{itemize}}
\def\benu{\begin{enumerate}}
\def\enu{\end{enumerate}}

\begin{document}
\coloredauthor{Raphael F. Ribeiro}
\coloredauthor{Joel Yuen-Zhou}
\affiliation{Department of Chemistry and Biochemistry, University of California San Diego, La Jolla, CA 92093}
\date{\today}
\title{Vibronic Ground-state Degeneracies and the Berry Phase: A Continuous Symmetry Perspective}



\begin{abstract}
We develop a geometric construction to prove the inevitability of the electronic ground-state (adiabatic) Berry phase for a class of Jahn-Teller models with maximal continuous symmetries and $N > 2$ intersecting electronic states. Given that vibronic ground-state degeneracy in JT models may be seen as a consequence of the electronic Berry phase, and that any JT problem may be obtained from the subset we investigate in this letter by symmetry-breaking, our arguments reveal the fundamental origin of the vibronic ground-state degeneracy of JT models.
\end{abstract}
\maketitle
The Jahn-Teller (JT) theorem \cite{jahn_stability_1937,jahn_stability_1938}
is a cornerstone of condensed matter and chemical physics; it enunciates
that adiabatic electronically degenerate states of symmetric nonlinear
molecules are unstable with respect to symmetry-breaking distortions
of the molecular geometry (unless the degeneracy is protected by time-reversal
symmetry). Given this statement, one might be tempted to loosely extrapolate
that molecular quantum state degeneracies are generally unstable.
This is, however, an incorrect conclusion: It is interesting that a
large class of JT models exhibit robust vibronic ground-state degeneracies
\cite{longuet-higgins_studies_1958,obrien_dynamic_1969-2,zwanziger_topological_1987,ham_berrys_1987,ham_role_1990,cullerne_jahn-teller_1994,de_los_rios_dynamical_1996,chancey1997jahn,bersuker_jahn-teller_2006,requist_molecular_2016-1,ryabinkin_geometric_2013}.
Thus, there is a counterintuitive flavor to the JT theorem: vibronic
degeneracies can be born at the expense of the breakdown of their
electronic counterparts \cite{bersuker_jahn-teller_2006,chancey1997jahn}. These degeneracies leave distinctive signatures in the chemical dynamics of JT systems which are sometimes immune to degeneracy-breaking perturbations \cite{ryabinkin_geometric_2013,joubert-doriol_molecular_2017}.
The goal of this letter is to explain the fundamental reason for the
emergence of degenerate vibronic ground-states in JT models. 

Vibronic ground-state degeneracy (VGSD) in JT models appears frequently
when linear vibronic couplings dominate \cite{bersuker_jahn-teller_2006, chancey1997jahn} (for a recent proposal of direct non-interferometric experimental verification of VGSD, see e.g.,\cite{englman_spectroscopic_2016,englman_non-interferometric_2016}),  although there are exceptions
\cite{moate_$mathithensuremathbigotimesmathith$_1996-1,bevilacqua2001nondegenerate,bersuker2001degenerate, lijnen_berry_2005}.
More specifically, there exists a particular class of JT models for
which VGSD is guaranteed to exist whenever the adiabatic approximation
(Born-Oppenheimer \cite{born_zur_1927} with inclusion of Berry phase
effects \cite{mead_determination_1979,berry_quantal_1984}) is valid
\cite{bersuker_jahn-teller_2006,chancey1997jahn,ribeiro_continuous_2017}.
These are the JT systems containing continuous symmetries \emph{and}
all possible couplings between JT active modes and a single electronically
degenerate multiplet (at the reference geometry for a description
of the JT effect, from now on denoted by JT \textit{center}) \cite{longuet-higgins_studies_1958,obrien_dynamic_1969-2,judd_lie_1974,judd_group_1982,pooler_jahn-teller_1977,pooler_continuous_1978,pooler_continuous_1980,ribeiro_continuous_2017},
the simplest and most famous example being the linear $E\otimes e$
model (we use the standard convention where the electronic irreducible
representation (irrep) is given by a capital letter and the vibrational
irrep is given by a lowercase) which displays an exotic SO(2) (circular)
symmetry in its potential energy surface \cite{longuet-higgins_studies_1958,bersuker_jahn-teller_2006}. The most complex spinless
example is the SO(5)-invariant model of the icosahedral JT problem
$H\otimes(g\oplus2h)$, which contains all possible JT active modes
associated with the electronic $H$ quintuplet \cite{pooler_continuous_1980,ceulemans_a._and_fowler_r._jahnteller_1990,de_los_rios_dynamical_1996,chancey1997jahn}.
On the other hand, the linearized $H\otimes h$ model has SO(3) symmetry
\cite{khlopin_jahn-teller_1978}, but it does not include the couplings
between $H$ and the remaining $g\oplus h$ active vibrations. In
this letter we focus only on the former class of models, which we
refer hereafter as JT systems with \textit{maximal} continuous symmetries
(MCSs) ($H\otimes h$ has SO(3) symmetry \cite{khlopin_jahn-teller_1978},
but inclusion of equally coupled $g\oplus h$ vibrations leads to
a JT Hamiltonian invariant under the action of SO(5) \cite{pooler_continuous_1980};
the latter is the maximal symmetry group available for a JT model
containing a single electronic $H$ multiplet \cite{pooler_continuous_1980,de_los_rios_dynamical_1996}).
Nevertheless, as we argue below, the results we obtain are expected
to be meaningful also in the presence of moderate symmetry-breaking perturbations, as continuous symmetry is not a \textit{necessary} condition for VGSD \cite{bersuker_jahn-teller_2006}).

We will discuss only JT models for which spin-orbit coupling can be
neglected, so we may take the system to be spinless and the time-reversal
operator $T$ to satisfy $T^{2}=1$. In all cases where it appears,
VGSD in JT models can be associated to a twisting of the lowest-energy
adiabatic electronic state as the molecular geometry traverses a loop
on a vibrational configuration space submanifold enclosing the jT center.
That is, a direct connection exists between the geometric phase \cite{berry_quantal_1984,mead_geometric_1992,bohm2003geometric}
and the exotic degeneracy in the molecular ground-state of a large
class of JT models \cite{zwanziger_topological_1987,ham_berrys_1987,ceulemans_berry_1991,cullerne_jahn-teller_1994,de_los_rios_dynamical_1996, chancey1997jahn,bersuker_jahn-teller_2006, requist_molecular_2016-1,ribeiro_continuous_2017,bohm2003geometric,ham_role_1990}.

In the limit where the ratio of the squared reduced (linear) vibronic constant
to the harmonic restoring force at the JT center is large, the extrema
of the electronic ground-state adiabatic potential energy surface
(APES) are located sufficiently far from the JT center \cite{bersuker_jahn-teller_2006}.
The energy gap between the electronic ground-state and any other state
in the considered multiplet is also generally large enough, and the
adiabatic approximation holds. In this case, VGSD in JT models with
MCSs arises from the following facts \cite{bersuker_jahn-teller_2006,chancey1997jahn}:
(i) Continuous symmetry implies the space of minima of the ground-state
APES is a continuous \textit{trough} \cite{pooler_continuous_1978,judd_jahnteller_1984,ceulemans_a._structure_1987,ribeiro_continuous_2017}.
(ii) If the choice is made that the electronic ground-state wavefunction
is real for any nuclear geometry, then it can only change by $\pm1$
when transported over a loop on the trough. (iii) Whenever this process
leads to a change in sign of the electronic ground-state wavefunction,
then, because the total vibronic wavefunction is single-valued, the
corresponding nuclear wavefunction must satisfy compensating antiperiodic
boundary conditions \cite{longuet-higgins_studies_1958,obrien_dynamic_1969-2,cullerne_jahn-teller_1994,de_los_rios_dynamical_1996,chancey1997jahn};
this turns out to be the case for JT models with MCSs. (iv) Motion
on the trough (pseudorotation) is equivalent to that of a free particle
on a sphere with antipodal points identified by an equivalence relation
\cite{obrien_dynamic_1969-2,ceulemans_a._structure_1987,de_los_rios_dynamical_1996,chancey1997jahn,ribeiro_continuous_2017},
and thus, the vibrational Schrodinger equation describing pseudorotation
is the same as that for a point particle constrained to move on a
sphere with appropriate boundary conditions. (v) The lowest energy wavefunction for a free particle on a spherical surface is symmetric under inversion; however, due to
item (iii), it cannot be the ground-state vibrational wavefunction
for JT models with MCSs. As a result, the molecular ground-state corresponds
to the lowest-energy multiplet of a particle on a sphere with wavefunctions
odd under inversion. (vi) This condition is satisfied by the vector
irreducible representation of the orthogonal group O($N$) (the symmetry
group of the sphere $S^{N-1}$). (viii) Finally, the vector irreps
of O($N$) have more than one real dimension for all $N>1$ (the O(2)
case is atypical, since the relevant representation to the corresponding
JT model is spanned by the time-reversal partners $e^{\pm i\phi/2}$;
this representation is only irreducible in the presence of time-reversal
symmetry) \cite{barut1986theory}; this result implies VGSD, where
the degeneracy is determined by the dimensionality of the vector irrep
of O($N$). The facts above have been verified on a case-by-case basis
for all JT models with continuous symmetries\cite{obrien_dynamic_1969-2,judd_jahnteller_1984,ham_berrys_1987,zwanziger_topological_1987,auerbach_vibrations_1994,cullerne_jahn-teller_1994,de_los_rios_dynamical_1996,chancey1997jahn}.
This list of items traces the origin of the aforementioned counterintuitive
feature of the JT problem: the electronic degeneracy, even though
lifted, leaves its signature in the resulting vibrational eigenspectrum.
By analogy to the Aharonov-Bohm effect \cite{aharonov_significance_1959}, as the nuclei circulate
the JT center, they nonlocally recognize its existence and inherit
a degeneracy themselves. 

Despite the existence of a variety of works on the Berry phase in JT and related models \cite{zwanziger_topological_1987, chancey1997jahn, ham_berrys_1987, de_los_rios_dynamical_1996, chancey_berrys_1988,ceulemans_berry_1991, apsel_berry_1992,cullerne_jahn-teller_1994,auerbach_vibrations_1994,schon_geometric_1995, manini_roberry_1998,bohm2003geometric,varandas_geometric_2004,lijnen_berry_2005,garcia-fernandez_lost_2006,bersuker_jahn-teller_2006, althorpe_effect_2008, zygelman_molecular_2017, abedi_correlated_2012, ryabinkin_geometric_2013, englman_spectroscopic_2016, requist_molecular_2016-1},
to the best of our knowledge, the following question has yet to be
answered: given that in a set of $N$ intersecting real electronic states,
such that some, but not all states change sign under a non-trivial
loop on the nuclear configuration space, why does the electronic ground-state
of JT models with MCSs (and in cases where this symmetry is only slightly
broken) \textit{always} has a nontrivial Berry phase (items (ii) and (iii) above)?
For example, consider the SO(3)-invariant version of the cubic JT
problem $T\otimes(e\oplus t_{2})$ \cite{obrien_dynamic_1969-2}.
Its APES trough has a constant electronic spectrum with a degenerate
branch including two excited electronic states, while the ground-state
is non-degenerate. As first verified by O'Brien \cite{obrien_dynamic_1969-2},
the real ground-state electronic wavefunction of this model is double-valued.
In other words, there exists an obstruction to the definition of a
continuous global real basis for the electronic ground-state \cite{chancey_berrys_1988,ceulemans_berry_1991}.
This obstruction implies the electronic ground-state has a Berry phase,
i.e., there exists loops on the vibrational configuration space,
which if traversed adiabatically lead to a change in the sign of
the electronic ground-state wavefunction \cite{longuet-higgins_intersection_1975,bohm2003geometric}.
The degenerate subspace orthogonal to the adiabatic ground-state is
spanned by two basis vectors, only one of which admits a nontrivial
Berry phase (see Eqs. 18 and 19). The more complex SO(4) and SO(5)
icosahedral models display the same features \cite{cullerne_jahn-teller_1994,de_los_rios_dynamical_1996,chancey1997jahn}. Notably, $E \otimes e$ is a special case as both the electronic ground- and excited-state display a non-trivial geometric phase \cite{bersuker_jahn-teller_2006}. Conversely, as mentioned above, for $N > 2$ there is \textit{a priori} no reason for the lowest energy electronic state to correspond to a non-trivial line bundle \cite{varandas_geometric_2004,varandas_geometrical_2010}. Thus, in this letter we aim attention at models with MCSs and $N > 2$. While the electronic ground-state Berry phase has been verified on a case-by-case basis for all of these models before, the steps required to demonstrate the existence of the Berry phase are algebraically lengthy even for relatively small $N$\cite{chancey1997jahn}. Additionally, prior arguments do not indicate the common origin of the Berry phase in all of these systems.
Because more realistic JT and other molecular systems with electronic
degeneracies may be understood to arise from symmetry breaking of
JT models with MCSs, an explanation for the inevitability of the electronic
ground-state Berry phase of the latter is also a foundation for an
understanding of the former.

It is the main objective of this letter to provide a simple answer to the
question raised in the previous paragraph. We aim to explain the basic
geometric reason for VGSD in JT systems
with MCSs. 
Importantly, JT models with MCSs
are minimal models for molecular conical intersections, so the
generic features we find here are also relevant for a wide variety
of problems in photochemical dynamics both in gas and condensed phases
\cite{atanasov_vibronic_2011,domcke2011conical,halasz_conical_2011,domcke_role_2012, gatti2014molecular} (for a recent review on the effects of the molecular Berry phase on nonadiabatic dynamics near conical intersection, see Ref. \cite{ryabinkin_geometric_2017-1}).

The  molecular Hamiltonian for a JT model with MCS can be written as:
\begin{equation} H = \frac{\mathbf{P}^2}{2} + \frac{\omega^2 \mathbf{Q}^2}{2} + H_{\text{JT}}(\mathbf{Q}), \end{equation}
where $\mathbf{Q}$ is the vector of nuclear displacements from the JT center $\mathbf{Q}=0$, $\mathbf{P}$ is the corresponding conjugate momentum, and $H_{\text{JT}}(\mathbf{Q})$ is the electronic JT Hamiltonian. The latter acts only on the family of $N-$dimensional electronic Hilbert spaces $\mathcal{H}_{el}(\mathbf{Q})$, and depends linearly on $\mathbf{Q}$. Thus, $H_{\text{JT}}(\mathbf{Q})$ may be written as
\begin{equation} 
H_{\text{JT}}(\mathbf{Q}) = F \sum_{k=1}^M Q_k V_k,
\end{equation}
where $F$ is the reduced vibronic coupling constant (from the Wigner-Eckart theorem), and $\{V_k\}$ are Clebsch-Gordan matrices depending on the choice of electronic \textit{diabatic} basis vectors $\ket{\psi_{k}}$, $k\in\{1,2,...,N\}$ forming a representation of the corresponding continuous group \cite{pooler_continuous_1978, pooler_continuous_1980}.  
A fundamental property of the JT models with MCSs is that they have a
continuous set of electronic ground-state minima $\mathcal{O}\subset\mathbb{R}^{M}$, where $\nabla_\mathbf{Q} \left[E_{\text{JT}}(\mathbf{Q}) + \omega^2 \mathbf{Q}^2/2 \right] = 0$
\cite{ceulemans_a._structure_1987,pooler_continuous_1978,pooler_continuous_1980,ribeiro_continuous_2017}.
In each case the electronic spectrum for any molecular geometry in
the trough $\mathcal{O}$ is given by \cite{longuet-higgins_studies_1958,obrien_dynamic_1969-2,ceulemans_so4_1989,chancey1997jahn,ribeiro_continuous_2017}
\begin{align}
\text{spec}[H_{\text{JT}}(\mathbf{Q})]=\{x(Q),x(Q),...,x(Q),-(N-1)x(Q)\},\nonumber \\
x(Q)>0,\mathbf{Q}\in\mathcal{O},\label{spec}
\end{align}
where $Q$ is the Euclidean length of the JT displacement vector $\mathbf{Q}$ (all
points of a given trough have the same value of $Q$), and $x(Q)$ is a real function of $Q$. For any $\mathbf{Q}\in\mathcal{O}$,
the JT Hamiltonian with spectrum given by Eq. \ref{spec}
may be rewritten as \cite{ribeiro_continuous_2017}

\begin{equation}
H_{\text{JT}}(\mathbf{Q})=\sum_{i=1}^{M}Q_{i}V_{i}=QU[\mathbf{\theta}(\mathbf{Q})]V_{M}U^{-1}[\mathbf{\theta}(\mathbf{Q})],
\end{equation}
where $QV_{M}$ is the diagonal matrix with entries determined by the electronic
spectrum for $\mathbf{Q}\in\mathcal{O}$ (in the order specified by
Eq. \ref{spec}), $\theta(\mathbf{Q})=(\theta_{1}(\mathbf{Q}),...,\theta_{N-1}(\mathbf{Q}))$
are SO$(N)$ parameters specifying a molecular geometry at $\mathcal{O}$
(if $\theta(\mathbf{Q})$ is multivalued, then a continuous local
choice of representative is assumed to have been made), and $U(\theta)$
(we will sometimes omit the dependence of $\theta$ on $\mathbf{Q}$
for notational simplicity) is the SO($N$) transformation of the electronic
Hilbert space at $\mathbf{Q}$ defined by 
\begin{equation}
Q[U(\theta)V_{M}U^{-1}(\theta)]=\sum_{i=1}^{M}\underbrace{(R^{-1}(\theta)\mathbf{Q}_{M}^T)_{i}}_{=Q_{i}}V_{i},
\end{equation}
where $R^{-1}(\theta)$ is the SO($N$)$\subset$SO($M$) vibrational
configuration space rotation (pseudorotation) which maps $\mathbf{Q}_{M}=Q\mathbf{e}_{M}$
into $\mathbf{Q}=\sum_{i=1}^{M}Q_{i}\mathbf{e}_{i}$, where the $\mathbf{e}_{i}$
are the unit vectors of the vibrational configuration space. For the sake of simplicity we chose $F=1$. Thus, we see that $\mathbf{Q}_M$ defines a reference JT distorted molecular structure for which $H_{\text{JT}}(\mathbf{Q})$ is already diagonal in the diabatic basis $\{\ket{\psi_k}\}$. Note
that $N\leq M$ for all JT models with MCSs \cite{pooler_continuous_1978, pooler_continuous_1980,judd_group_1982,ribeiro_continuous_2017}. As an example of the above definitions, consider the case of $E \otimes e$. Since only two electronic states are retained in this model, it follows that $N = 2$, and the diabatic basis may be written as $\{\ket{\psi_1},\ket{\psi_2}\}$. Let $\mathbf{\sigma} = (\sigma_x,\sigma_z)$ denote a matrix vector (each entry corresponds to a Pauli matrix), and $\mathbf{Q}  = (Q_x, Q_z) = Q \text{sin}\theta \mathbf{e}_x+ Q \text{cos}\theta \mathbf{e}_z$ [where $\theta = \text{tan}^{-1}(Q_x/Q_z)$] be the JT displacement from the maximally symmetric structure at $\mathbf{Q}=0$. In this case, the electronic Hamiltonian can be written as
\small
\begin{align}  
& H_{\text{JT}}^{E\otimes e} (\mathbf{Q}) = \mathbf{Q}\cdot \sigma^T = 
Q \text{sin} (\theta) \sigma_x+Q\text{cos}(\theta) \sigma_z , \\ & H_{\text{JT}}^{E\otimes e} (\mathbf{Q})  = Qe^{-i\sigma_y \theta/2}\sigma_z e^{i \sigma_y \theta/2}, \\
& H_{\text{JT}}^{E\otimes e} (\mathbf{Q}) = QU[\theta(\mathbf{Q})]\sigma_zU^{-1}[\theta(\mathbf{Q})],~\text{where}~ \\ & U[\theta(\mathbf{Q})]  = e^{-i \sigma_y \theta(\mathbf{Q})/2}.\end{align}
\normalsize
In the notation of Eq. 4, it follows that for $E \otimes e$,  $M = 2$ and $V_2 = \sigma_z$. Hence, for a given $Q$, the molecular structure with vanishing diabatic couplings is given by $\theta = 0$. We obtain the relationship expressed by Eq. 5 by setting $\mathbf{e}_2 = \mathbf{e}_z$ and $\mathbf{Q}_2 = Q(0,1)$, such that
\begin{equation} H_{\text{JT}}^{E\otimes e} (\mathbf{Q}) = Q U(\theta) V_2 U^{-1}(\theta) =  \sum_{i=1}^2 \left[R^{-1}(\theta)\mathbf{Q}_2^T \right]_i \sigma_i, \end{equation}
where we employed the relationship
\begin{equation} R^{-1}(\theta) \mathbf{e}_z^T =  \begin{pmatrix}
\text{cos}\theta & \text{sin}\theta \\
-\text{sin}\theta & \text{cos}\theta \end{pmatrix}\cdot \mathbf{e}_z^T  = \left( \text{sin}\theta, \text{cos}\theta \right).\end{equation}
Eqs. $3-5$ generalize the $E \otimes e$ construction and display the basic property of JT models with MCSs: a
change of basis of the electronic Hilbert space preserving its real
structure (e.g., an SO($N)$ transformation) leads to an electronic
Hamiltonian matrix that can also be obtained by a rotation of the
vibrational configuration space \cite{pooler_continuous_1978, pooler_continuous_1980,judd_group_1982,ribeiro_continuous_2017}.
However, note that only $N-1$ of the $N(N-1)/2$ degrees of freedom
of SO($N$) are required to identify a point of the ground-state trough
of JT models with MCSs. This is a consequence of the spectrum given
by Eq. 1. In particular, the subgroup SO$(N-1)\subset$ SO$(N)$ that
acts non-trivially only on the electronically degenerate subspace
commutes with $H_{\text{JT}}(\mathbf{Q})$. Therefore, its corresponding
action on the vibrational configuration space is trivial and gives
rise to no additional molecular structures with the electronic spectrum
given by Eq. 1.

Let the eigenstate of $H_{\text{JT}}(\mathbf{Q}_{M})$ with lowest
eigenvalue be written as $\ket{\phi_{0}(\mathbf{Q}_{M})}=\ket{\psi_{N}}$,
(from Eqs. 4 and 5, it follows that $U[\theta(\mathbf{Q}_{M})]=U(0)=R(0)=1$).
For any $\mathbf{Q}\in\mathcal{O}$ related to $\mathbf{Q}_{M}$ by
a pseudorotation, i.e., $\mathbf{Q}=R^{-1}(\theta)\mathbf{Q}_{M}$,
a normalized electronic ground-state wavefunction of $H_{\text{JT}}(\mathbf{Q})$
is 
\begin{align}
 & \ket{\phi_{0}(\mathbf{Q})}=U(\theta)\ket{\phi_{0}(\mathbf{Q}_{M})}=U(\theta)\ket{\psi_{N}},\nonumber \\
 & \ket{\phi_{0}(\mathbf{Q})}=\sum_{i=1}^{N}c_{0i}[\theta(\mathbf{Q})]\ket{\psi_{i}},c_{0i}\in\mathbb{R}.\label{gsad}
\end{align}
The excited states span the hyperplane $\mathcal{S}_{0}^{\perp}(\mathbf{Q})$
of the electronic Hilbert space $\mathcal{H}_{el}(\mathbf{Q})$ at
$\mathbf{Q}\in\mathcal{O}$ that is perpendicular to the line defined
by the adiabatic electronic ground-state $\mathcal{L}_{0}(\mathbf{Q})=\text{span}\{\ket{\phi_{0}(\mathbf{Q})}\}$.
Because the ground-state of $H_{\text{JT}}(\mathbf{Q})$ is gapped
we may decompose $\mathcal{H}_{el}(\mathbf{Q})$ uniquely for any
$\mathbf{Q}\in\mathcal{O}$ into the direct sum: 
\begin{equation}
\mathcal{H}_{el}(\mathbf{Q})=\mathcal{L}_{0}(\mathbf{Q})\oplus\mathcal{S}_{0}^{\perp}(\mathbf{Q}).
\end{equation}

In addition, since $T^{2}=1$, we take $\mathcal{H}_{el}(\mathbf{Q})$
to be a real vector space. Thus, a normalized basis for $\mathcal{L}_{0}(\mathbf{Q})$
is given by Eq. \ref{gsad}. The only permissible orthogonal basis
transformations of $\mathcal{L}_{0}(\mathbf{Q})$ are given by multiplication
by $O(1)=\pm1$. Conversely, any choice of basis for $\mathcal{S}_{0}^{\perp}(\mathbf{Q})$
can be redefined by orthogonal transformations belonging to O$(N-1)$.

For every $\mathbf{Q}\in\mathcal{O}$ we can define a sphere $S^{N-1}(\mathbf{Q})$
immersed in $\mathbb{R}^{N}$. Then, because $\ket{\phi_{0}(\mathbf{Q})}=\sum_{i=1}^{N}c_{0i}[\mathbf{Q}]\ket{\psi_{i}}$
belongs to a line, we can represent it as the outwards normal vector of $S^{N-1}(\mathbf{Q})$
at the point with coordinates $c_{0}[\mathbf{Q}]=(c_{01}[\mathbf{Q}],...,c_{0N}[\mathbf{Q}])$.
The hyperplane $\mathcal{S}_{0}^{\perp}(\mathbf{Q})$ can be mapped
onto the tangent space of $S^{N-1}(\mathbf{Q})$ at the point $c_0[\mathbf{Q}]$,
i.e., there exists a map $\mathcal{S}_{0}^{\perp}(\mathbf{Q})\mapsto T_{c_0[\mathbf{Q}]}S^{N-1}(\mathbf{Q})$.
The mappings described above can be locally given by hyperspherical
unit vectors (e.g., if $N=3$, then the tangent space of $c_0[\mathbf{Q}]\in S^{2}(\mathbf{Q})$
can be taken as the span of the polar and azimuthal vectors $\mathbf{e}_{\theta}(\mathbf{Q})$
and $\mathbf{e}_{\phi}(\mathbf{Q})$, while the normal vector field
is $\mathbf{e}_{r}(\mathbf{Q})$). 

Adiabatic transport of the ground-state $\ket{\phi_{0}(\mathbf{Q})}$
over the trough $\mathcal{O}$ is implemented by defining a curve
$\text{\ensuremath{\gamma}:}~[0,1]\rightarrow\mathcal{O}$ along which
$\ket{\phi_{0}(\mathbf{Q})}$ is parallel-transported according to
the connection defined by \cite{berry_quantal_1984,ceulemans_berry_1991,bohm2003geometric},
\begin{equation}
\braket{\phi_{0}(\mathbf{Q})|\mathrm{d}\phi_{0}(\mathbf{Q})}=0.\label{adcon}
\end{equation}
This condition is necessarily satisfied by any choice of real local
section of $\mathcal{L}_{0}(\mathbf{Q})$ (recall there exists two (normalized)
possibilities $\pm c_{0}[\mathbf{Q}]$ for the electronic ground-state
at a given $\mathbf{Q}\in\mathcal{O}$; a local section is a continuous choice
of either one of those for some open subset of $\mathcal{O}$). 

Now consider an adiabatic loop starting at arbitrary $\mathbf{Q}_{0}\in\mathcal{O}$
\begin{align}
\text{\ensuremath{\gamma_{L}}:}~ & [0,1]\rightarrow\mathcal{O},\nonumber \\
 & t\mapsto\gamma_{L}(t),~\gamma_{L}(0)=\gamma_{L}(1)=\mathbf{Q}_{0}.
\end{align}
As $t$ varies between 0 and 1, $\mathbf{Q}$ traverses the closed
path $\gamma_{L}(t)\in\mathcal{O}$ and $\ket{\phi_{0}[\gamma_{L}(t)]}$
is parallel-transported according to the adiabatic connection (Eq. \ref{adcon}). Its
associated normal vector traces a path on the space $\mathcal{B}$
defined by the disjoint union of the spheres $S^{N-1}(\mathbf{Q})$
attached to each $\mathbf{Q}\in\mathcal{O}$, 
\begin{equation}
\mathcal{B}\equiv\bigsqcup_{\mathbf{Q}\in\mathcal{O}}S^{N-1}(\mathbf{Q}).\label{bundle}
\end{equation}

Intuitively, parallel transport ensures that given an initial vector
$\ket{\phi_{0}(\mathbf{Q}_{0})}$ and a continuous path $\gamma_{L}$
in $\mathcal{O}$, there is a uniquely defined continuous curve $\ket{\phi_{0}[\gamma_{L}(t)]}$.
If adiabatic transport along $\gamma_{L}(t)$ corresponds to an open path on $\mathcal{B}$,
then it must take $c_{0}(\mathbf{Q}_{0})$ at $t=0$ into $-c_{0}(\mathbf{Q}_{0})$
at $t=1$. In this case, while the nuclei undergo a loop in the space
of allowed JT distortions, the normal vector corresponding to the
electronic ground-state is mapped into its antipode; thus, a Berry
phase ensues. Continuous loops $\gamma_{L}(t)\in\mathcal{O}$ satisfying
the preceding conditions always exist, as the electronic ground-state
trough $\mathcal{O}$ is topologically equivalent to the real projective
space $\mathbb{R}P^{N-1}$ \cite{ceulemans_a._structure_1987, de_los_rios_dynamical_1996,ribeiro_continuous_2017} (for a detailed discussion of this point, see Secs.III.B.1 and III.B.2 of Ref. \cite{ribeiro_continuous_2017}).
$\mathbb{R}P^{N-1}$ has loops that are lifted to open paths connecting
antipodal points of $S^{N-1}$ \cite{ceulemans_a._structure_1987,ribeiro_continuous_2017,nakahara2003geometry, lee2010introduction}.
These features are crucial elements of our proof. The topological
equivalence between $\mathcal{O}$ and $\mathbb{R}P^{N-1}$ can be
simply restated as there being a continuous bijection between molecular geometries
$\mathbf{Q}\in\mathcal{O}$ and \textit{real} pure-state projection operators $|\eta\rangle\langle\eta|$
with $|\eta\rangle=\sum_{i=1}^{N}d_{i}|\psi_{i}\rangle,~d_i \in \mathbb{R}$. 

More formally, the argument just given may be rephrased in the following
way: $\mathcal{O}\cong\mathbb{R}P^{N-1}$ \cite{ceulemans_a._structure_1987, de_los_rios_dynamical_1996, chancey1997jahn, ribeiro_continuous_2017} implies the existence of a continuous bijective map (with continuous
inverse)

\begin{align}
\text{\ensuremath{\Phi}:}~ & \mathcal{O}\rightarrow\mathbb{R}P^{N-1},\nonumber \\
 & \mathbf{Q}\mapsto\ket{\phi_{0}(\mathbf{Q})}\bra{\phi_{0}(\mathbf{Q})}.
\end{align}
As a result, the equivalence classes of loops (containing all closed
paths which can be deformed continuously into each other) of $\mathbb{R}P^{N-1}$
and the space of ground-state minima $\mathcal{O}$ are equal \cite{nakahara2003geometry,lee2010introduction}.
The case where $N=2$ corresponds to the thoroughly investigated SO(2)-invariant
linear $E\otimes e$ system \cite{longuet-higgins_studies_1958,zwanziger_topological_1987,ceulemans_berry_1991,bersuker_jahn-teller_2006},
the only JT model with a spherical trough (as $\mathbb{R}P^{1}\cong S^{1}$,
but $\mathbb{R}P^{N-1}\neq S^{N-1}$ when $N>2$ \cite{nakahara2003geometry,lee2010introduction}). From now on we assume $N>2$.

The non-trivial class $[\gamma_{L}(t)]$ of $\mathbb{R}P^{N-1}$ loops
can be lifted via the adiabatic connection (and a choice of local
section for the initial point on $S^{N-1}(\mathbf{Q})$, e.g., $c_{0}[\mathbf{Q}]$
representing $\ket{\phi_{0}(\mathbf{Q})}$) to a class of open paths
on $\mathcal{B}$: let $(\mathbf{Q},p)$ with $p \in S^{N-1}(\mathbf{Q})$,
denote local coordinates for the bundle $\mathcal{B}$ defined by
Eq. \ref{bundle}. Then, the lift to $\mathcal{B}$ of a non-trivial loop $\gamma_{L}(t) \in [\gamma_{L}(t)]$ is defined by the curve $\text{\ensuremath{\tilde{\gamma}_{L}}:}~[0,1]\rightarrow\mathcal{B}$
connecting the points $(\mathbf{Q}_{0},\pm c_{0}[\mathbf{Q}_{0}])$
\cite{nakahara2003geometry,lee2010introduction}, i.e., 
\begin{align}
 & \tilde{\gamma}_{L}(t)=(\gamma_{L}(t),C[\gamma_{L}(t)]),~t\in[0,1],\nonumber \\
 & C[\gamma_{L}(0)]=(c_{01}[\mathbf{Q}_{0}],...,c_{0N}[\mathbf{Q}_{0}])=\ket{\phi_{0}(\mathbf{Q}_{0})},\nonumber \\
 & C[\gamma_{L}(1)]=-C[\gamma_{L}(0)]=-\ket{\phi_{0}(\mathbf{Q}_{0})},\label{loop}
\end{align}
where $C[\gamma_{L}(t)]$ is the vector obtained by parallel transport
of $\ket{\phi_{0}(\mathbf{Q}_{0})}$ along $\gamma_{L}(t)$ with the
adiabatic connection (Eq. \ref{adcon}). The geometric phase of the
electronic ground-state for \textit{any} loop $\gamma'(t)\in\mathcal{O}$
with lift $\tilde{\gamma}'(t)\in\mathcal{B}$ is given by \cite{berry_quantal_1984, ceulemans_berry_1991, mead_geometric_1992}
\begin{equation}
\Gamma_{\gamma'}^{0}=\braket{C[\gamma'(0)]|C[\gamma'(1)]}.
\end{equation}
Given that $\ket{\phi_{0}(\mathbf{Q})}$ is a normal vector of $S^{N-1}(\mathbf{Q})$,
and a non-trivial loop $\gamma_{L}$ of $\mathbb{R}P^{N-1}$ is lifted
into an open path on the family (bundle) of spheres attached to each $\mathbf{Q}\in\mathcal{O}$,
it follows that $\gamma_{L}$ provides a mapping of the normal vector
at $c_{0}[\mathbf{Q}]\in S^{N-1}(\mathbf{Q})$ into that at $-c_{0}[\mathbf{Q}]\in S^{N-1}(\mathbf{Q})$
which in turn implies $\Gamma_{\gamma_{L}}^{0}=-1$. Figure 1 illustrates
this result (for visualization purposes the representative lift $\tilde{\gamma}_{L}(t)$
is drawn on a single sphere).

\begin{figure}
\includegraphics[width=0.5\textwidth]{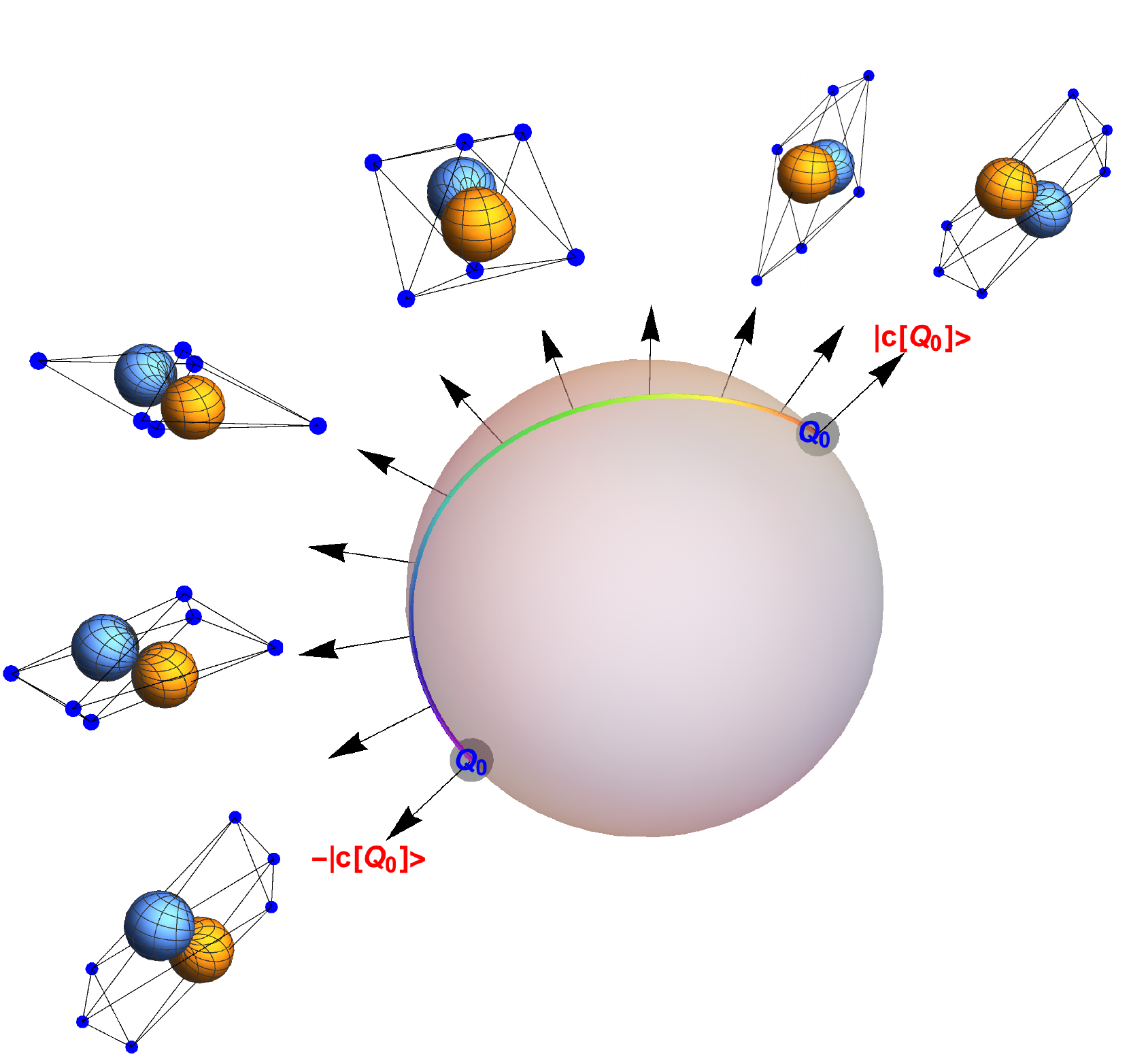}
\caption{A visual representation of the lift $\tilde{\gamma}_{L}(t)$ (Eq.
\ref{loop}) of a non-trivial loop on the ground-state minimal energy
trough $\mathcal{O}$ based at $\mathbf{Q}_{0}$, to the sphere bundle
$\mathcal{B}$ for the SO(3)-invariant cubic model $T_{1}\otimes(e\oplus t_{2})$
\cite{obrien_dynamic_1969-2} (see Eqs. \ref{gsex} and \ref{exex}).
In this figure the path on the family of spheres parametrized by $\mathbf{Q}$
is represented on a single sphere for simplicity. The arrows correspond
to a choice of local section for the normal vectors representing $\mathcal{L}_{0}(\mathbf{Q})$.
The nontrivial $-1$ Berry phase of this model is indicated by the fact
that the normal vector at $\mathbf{Q}_{0}$, $n(\mathbf{Q}_{0})$
is mapped into $-n(\mathbf{Q}_{0})$ under adiabatic transport over
a loop on $\mathcal{O}$.}
\label{spherebuncurve} 
\end{figure}

Our argument also implies that the number of linearly independent
states which change sign under adiabatic transport on $\mathcal{O}$
is equal to the number of hyperspherical unit vectors of $S^{N-1}$
which have their sign changed under inversion of the sphere. For example,
in the SO(3)-invariant model of $T_{1}\otimes(e\oplus t_{2})$ \cite{obrien_dynamic_1969-2}
the adiabatic ground-state is given by 
\begin{equation}
\mathbf{e}_{r}\equiv\ket{\phi_{0}(\theta,\phi)}=\text{sin}\theta\text{cos}\phi\ket{\psi_{1}}+\text{sin}\theta\text{sin}\phi\ket{\psi_{2}}+\text{cos}\theta\ket{\psi_{3}},\label{gsex}
\end{equation}
where $\ket{\psi_{i}},i=1,2,3$, are electronic basis vectors spanning
the $T_{1}$ irrep of the octahedral group, and for $\mathbf{Q}\in\mathcal{O}$
it follows that $\mathbf{Q}=\mathbf{Q}(\theta,\phi)=\mathbf{Q}(\pi-\theta,\phi+\pi)$
where $\theta\in[0,\pi],\phi\in[0,2\pi)$ \cite{obrien_dynamic_1969-2}.
In fact, $\ket{\phi_{0}(\theta,\phi)}=-\ket{\phi_{0}(\pi-\theta,\phi+\pi)}$.
The family of degenerate subspaces orthogonal to $\mathbf{e}_{r}(\theta,\phi)$
is generated by the basis vectors $\mathbf{e}_{\theta}(\theta,\phi)\equiv\ket{\phi_{1}(\theta,\phi)}$,
and $\mathbf{e}_{\phi}(\theta,\phi)\equiv\ket{\phi_{2}(\theta,\phi)}$,
where
\begin{align}
 & \ket{\phi_{1}(\theta,\phi)}=\text{cos}\theta\text{cos}\phi\ket{\psi_{1}}+\text{cos}\theta\text{sin}\phi\ket{\psi_{2}}-\text{sin}\theta\ket{\psi_{3}},\nonumber \\
 & \ket{\phi_{2}(\theta,\phi)}=-\text{sin}\phi\ket{\psi_{1}}+\text{cos}\phi\ket{\psi_{2}}.\label{exex}
\end{align}
Note that $\mathbf{e}_{\theta}[\mathbf{Q}]$ is fixed, but $\mathbf{e}_{\phi}[\mathbf{Q}]$
has its direction inverted when $(\theta,\phi)\mapsto(\pi-\theta,\pi+\phi)$.
Similar verifications may be performed for the SO(4) and SO(5) icosahedral
JT models \cite{varandas_geometric_2004,varandas_geometrical_2010}.
In every case, only the normal vector corresponding to the ground-state
and one of the spherical basis vectors of $\mathcal{S}_{0}^{\perp}(\mathbf{Q})$
acquire a non-trivial phase when $\mathbf{Q}$ traverses a loop
on $\mathcal{O}$. This can be understood by considering the behavior
of hyperspherical basis vectors under inversion of $S^{N-1}$. Previous
work by Varandas and Xu \cite{varandas_geometric_2004} provided
the possible sign changes of the electronic adiabatic states of JT
models (see also Ref.\cite{varandas_geometrical_2010}) by explicitly
constructing the higher-dimensional counterparts of Eqs. \ref{gsex}
and \ref{exex}, though their considerations did not uncover the fundamental
reason the lowest-energy state of JT models with MCSs always admits a nontrivial Berry phase.

At first sight our construction may be perceived as severely restricted
by the conditions that (a) the molecular Hamiltonian is totally symmetric
under the action of a continuous group on the electronic and vibrational
degrees of freedom and (b) the ground-state trough $\mathcal{O}$
has the electronic spectrum described by Eq. 1. However, higher-order vibronic perturbations
which remove these constraints may only change the Berry phase if
they induce degeneracies between the adiabatic ground-state $\ket{\phi_{0}(\mathbf{Q})}$
and any other electronic state in regions relevant to nuclear dynamics
at low energies. This will happen e.g., if sufficiently strong quadratic
vibronic couplings are introduced \cite{zwanziger_topological_1987,koizumi_multiconical_1999,koizumi_multiple_2000}. Alternatively, any external perturbation (e.g., due to a static electric field) which couples the degenerate states will lift their degeneracy \cite{bersuker_jahn-teller_2006}. In other words, while generic perturbations typically break MCSs and lead to a discrete set of ground-state minima (as opposed to a continuous trough) \cite{zwanziger_topological_1987,koizumi_multiconical_1999,koizumi_multiple_2000,bersuker_jahn-teller_2006} the Berry phase may still persist. Therefore, while the existence of a ground-state trough is significant to our proof, it is \textit{not} a necessary condition for the existence of VGSD.

In summary, the reason that VGSD exists in spinless JT models with
MCSs is that the adiabatic electronic ground-state corresponding to
every geometry $\mathbf{Q}$ in the space $\mathcal{O}$ of JT APES
minima can be canonically mapped to a normal vector of a sphere attached
to each $\mathbf{Q}$, while its orthogonal subspace can be mapped
onto the tangent space at the same point. Under adiabatic transport
over a non-trivial loop on $\mathcal{O}$ the spherical normal vector
corresponding to the electronic ground-state has its direction reversed,
thereby giving a Berry phase, and requiring antiperiodic boundary
conditions to be satisfied by the nuclear wavefunction, for the total
vibronic ground-state to be single-valued. It is crucial for the existence
of a Berry phase that the aforementioned non-trivial paths exist and
are relevant for the dynamics of the physical system at low energies.
These pre-requisites are ensured here by the topological equivalence
between the nuclear trough $\mathcal{O}$ and the electronic projective
space $\mathbb{R}P^{N-1}$ \cite{ceulemans_a._structure_1987,ribeiro_continuous_2017},
and by the fact that the Born-Oppenheimer potential energy is constant
in $\mathcal{O}$. In other words, our construction relied on two
key features of JT models with MCSs: (i) the topological equivalence
between the JT trough $\mathcal{O}$ and the real projective Hilbert
space, and (ii) the existence of a uniquely defined electronic ground-state
line $\mathcal{L}_{0}(\mathbf{Q})$ and excited degenerate subspace
$\mathcal{S}_{0}^{\perp}(\mathbf{Q})$ perpendicular to $\mathcal{L}_{0}(\mathbf{Q})$
at all $\mathbf{Q}\in\mathcal{O}$. Although these conditions are
ideal, as explained, VGSD is protected as long as perturbations breaking
the MCS of the studied models are not strong enough that intersections
between the electronic ground-state and the remaining states emerge
in low-energy regions of the ground-state APES. Therefore, we believe
the presented construction provides the fundamental reason for the
prevalence of VGSD in a large class of JT models. It would be interesting
to understand whether these topological degeneracies can emerge in
other contexts, such as optical or mechanical systems.

\emph{Acknowledgments.\textendash{}} 
Both authors acknowledge support from NSF CAREER award CHE:1654732
and generous UCSD startup funds.

\end{document}